\documentclass{article}

\usepackage{arxiv}

\usepackage[utf8]{inputenc} 
\usepackage[T1]{fontenc}    
\usepackage{hyperref}       
\usepackage{url}            
\usepackage{booktabs}       
\usepackage{amsfonts}       
\usepackage{nicefrac}       
\usepackage{microtype}      
\usepackage{lipsum}
\usepackage{longtable}
\usepackage{authblk}
\usepackage{graphicx}
\usepackage{multirow}
\usepackage{subcaption}
\usepackage{caption}
\graphicspath{ {./images/}}

\title{Data Curation and Quality Assurance for Machine Learning-based Cyber Intrusion Detection}

\author[1]{Haihua Chen}
\author[1]{Ngan Tran}
\author[1]{Anand Sagar Thumati}
\author[2]{Jay Bhuyan}
\author[\space\space 1]{Junhua Ding\thanks{Corresponding author. E-mail address: junhua.ding@unt.edu.}}
\affil[1]{Department of Information Science, University of North Texas, Denton, TX, 76203, USA}
\affil[2]{Department of Computer Science, Tuskegee University, Tuskegee, AL, 36088, USA}

\begin{document}
\maketitle
\begin{abstract}
Intrusion detection is an essential task in the cyber threat environment. Machine learning and deep learning techniques have been applied for intrusion detection. However, most of the existing research focuses on the model work but ignores the fact that poor data quality has a direct impact on the performance of a machine learning system. More attention should be paid to the data work when building a machine learning-based intrusion detection system. This article first summarizes existing machine learning-based intrusion detection systems and the datasets used for building these systems. Then the data preparation workflow and quality requirements for intrusion detection are discussed. To figure out how data and models affect the machine learning performance, we conducted experiments on 11 HIDS datasets using seven machine learning models and three deep learning models. The experimental results show that BERT and GPT were the best algorithms for HIDS on all of the datasets. However, the performance on different datasets vary, indicating the differences between the data quality of these datasets. We then evaluate the data quality of the 11 datasets based on quality dimensions proposed in this paper to determine the best characteristics that a HIDS dataset should possess in order to yield the best possible result. This research initiates a data quality perspective for researchers and practitioners to improve the performance of machine learning-based intrusion detection.
\end{abstract}


\section{Introduction}
The increasing usage of digital devices in a cyber-physical system (CPS) has enhanced the efficiency of operating systems, but has also led to vulnerability from cyber-attacks. Cyber assaults on process control and the monitoring of these intelligent systems could lead to a significant control failure \cite{ayodeji2020new} and huge economic losses. This makes cyber-security a major concern due to the high level of attacks on networks and systems for CPS \cite{chawla2018host}. Therefore, building an intrusion detection system (IDS) to predict and respond to assaults on a CPS, has become an essential task among the software engineering community. However, it is very challenging because of the range of novelty involved in cyber-attacks \cite{ayodeji2020new}. Recently, machine learning and deep learning have been applied for intrusion detection at both the operation system level (a host-based intrusion detection system, called HIDS) and the network level (a network-based intrusion detection system, called NIDS). As indicated by Google Research \cite{10.1145/3411764.3445518}, both the models/algorithms and the quality of the data greatly impact the performance of machine learning systems. The computing rule of ``garbage in, garbage out'' is still applicable to machine learning \cite{GenderShade:Buolamwini18a} and the lack of high quality training data becomes a barrier for building high performance machine learning systems \cite{9282280}. Therefore, we should not only optimize the models but also systematically evaluate and ensure the data quality to improve performance for intrusion detection.  

However, current studies on machine learning-based intrusion detection only focus on model construction and optimization. For example, Sahu et al. compared the performance of various linear and non-linear classifiers for NIDS using the KDD dataset \cite{sahu2020data}, while Al-Maksousy compared deep neural networks (DNN) and various traditional machine learning (ML) models for NIDS on the same dataset, finding that DNN outperformed ML models in terms of accuracy, running time, and false positive rates \cite{al2018nids}. Hu et al. proposed an incremental HMM training framework that incorporates a simple data pre-processing method for identifying and removing similar sub-sequences of system calls for HIDS \cite{hu2009simple}. This training strategy has been widely applied since it can save on the training cost (especially on large data) without noticeable degradation of intrusion detection performance \cite{hu2009simple}. Convolutional neural network (CNN) and recurrent neural network (RNN) have also been used in HIDS \cite{tran2017approach, chawla2018host}. Recently, Liu and Lang conducted a comprehensive survey on machine learning and deep learning methods for IDS \cite{liu2019machine}. Nevertheless, very few research studies have paid attention to data requirements, data quality issues, and data quality assurance for IDS. 

The input of machine learning-based intrusion detection is a collection of data instances, each of which is represented by only one feature or a group of features. The features can be binary, categorical, or continuous. Data instances can be related to each other, such as sequence data, spatial data, and graph data. Each instance can be labeled as normal class or anomaly class, or is not labeled. The data requirements for different machine learning techniques vary. A training dataset should be labeled with normal and anomaly classes for supervised learning. The model is trained on the labeled data, then any unseen data is input to the trained model to determine to which class it belongs \cite{chandola2009anomaly}. Semi-supervised learning assumes labeled instances only to include either normal class or anomaly class, then the trained model is used to identify anomalies in the test data \cite{chandola2009anomaly}. Noisy labels, insufficient labeled data, and class imbalance are the most common data quality issues for supervised and semi-supervised learning-based IDS. Surprisingly, unsupervised learning does not require labeled data \cite{chandola2009anomaly}. However, unsupervised learning techniques usually assume that normal instances are far more frequent than anomalies in the test data. Under this assumption, semi-supervised learning can be incorporated with unsupervised learning by taking a sample of the unlabeled dataset as training data to improve the robustness of the model. Other data quality issues, such as inconsistency, duplication, incompleteness, incomprehension, no variety, or imprecise timestamps, might also exist in the data input of all the machine learning models. 

Instead of optimizing the machine learning-based IDS from the model/algorithm perspective, this paper targets the data-centric IDS and discusses how to systematically evaluate and ensure the data quality to improve performance for intrusion detection. We will answer a few important questions regarding the data quality to the performance of machine learning-based IDS: 

\begin{itemize}
    \item How should an appropriate dataset be prepared for intrusion detection in a specific scenario? What are the data requirements? We investigate and compare the existing datasets used for different intrusion detection tasks to answer these questions.
    \item How should data quality be evaluated? How is it possible to judge whether the data quality or the machine learning model has a major effect on the intrusion detection performance? How does data quality affect the intrusion detection performance? We conduct a case study on a HIDS, which aims to detect user anomalous behaviors in an operating system based on system call sequences. Ten different machine learning and deep learning techniques are implemented for the intrusion detection. A comprehensive analysis on the experiments is then performed to answer the second question. 
\end{itemize}
 
To the best of our knowledge, this is the first study which explores intrusion detection from a data-centric rather than a model-centric perspective. It will benefit IDS researchers and practitioners with new insights on improving intrusion detection performance by enhancing the data quality. The rest of the paper is structured as follows: Section 2 presents the literature review related to machine learning-based IDS and data quality. Section 3 discusses the data preparation and quality requirements for intrusion detection. Section 4 discusses the experiments, experimental results, analysis, and data quality assessment. Section 5 concludes the paper and discusses future work. The code and datasets used in this research are available on GitHub \footnote{\url{https://github.com/anandsagarthumati9848/HIDS}}.
         
\section{Related work}
\label{sec:headings}

\subsection{Machine learning-based intrusion detection systems}

Intrusion detection aims to detect malicious activities or intrusions (break-ins, penetrations, and other forms of computer abuse) in a computer-related system (operating system or network system) \cite{chandola2009anomaly}. An intrusion acts differently than the normal behavior of the system and, hence, the techniques used for anomaly detection can also be used for intrusion detection. 

Machine learning techniques used for intrusion detection can be divided into supervised learning and unsupervised learning. Whether the labeling of data is sufficient or not becomes the key criteria for selecting a machine learning technique. However, the detection performance of unsupervised learning methods is usually inferior to those of supervised learning methods \cite{liu2019machine}. Meanwhile, due to issues with how the data for intrusion detection typically flows (in a streaming fashion) and the data imbalance caused by low false alarm rates, the usage of machine learning techniques in intrusion detection is more challenging than other anomaly detection applications. Table \ref{table:mlmodels-new} summarizes the machine learning algorithms used for HIDS and NIDS in the last five years. 

\begin{center}
\begin{longtable}{p{2.5cm}p{5.0cm}p{2cm}p{2cm}p{1.2cm}p{1.3cm}}
\caption{ \centering Machine learning techniques for IDS. Full names and abbreviations of the models are introduced in the Appendix \ref{Appendix:A}.} \\
\hline
\textbf{Technique used}& \textbf{Model}& \textbf{Datasets}& \textbf{HIDS/NIDS}& \textbf{Year}& \textbf{Reference}\\
\hline
Supervised & KNN, SR & KDD-Cup99  & NIDS & 2017 & \cite{zhao2017dimension}\\
\hline
Supervised & LR, RF
 & UNSW & NIDS & 2017 & \cite{salman2017machine}\\
\hline
Supervised & DNN, imbalanced network traffic, RF, VAE & CIDDS-001 & NIDS & 2018 & \cite{abdulhammed2018deep}\\
\hline
Supervised & DNN & NSL-KDD & NIDS & 2018 & \cite{behera2018deep}\\
\hline
Supervised & DNN  
 & NSL-KDD & NIDS & 2018 & \cite{naseer2018enhanced}\\
\hline
Supervised & LR, NB, KNN, DT, AdaBoost, RF, CNN, CNN-LSTM, LSTM, GRU, SimpleRNN, DNN & CICIDS-2017, UNSW-NB15, ICS cyberattack  & NIDS & 2018 & \cite{elmrabit2020evaluation}\\
\hline
Unsupervised & Metric learning + clustering + SVM & Kyoto 2006, NSL-KDD & NIDS & 2019 & \cite{aliakbarisani2019data}\\
\hline
Supervised & NB, AODE, RBFN, MLP, J48 DT  
 & UNSW-NB15 & NIDS & 2019 & \cite{nawir2019effective}\\
\hline
Supervised & Pruned exact linear time, quantile regression forests
 & NetFlow data & NIDS & 2020 & \cite{evangelou2020anomaly}\\
\hline
Unsupervised & Autoencoder, IF, KNN, K-Means, SCH, SVM
 & NSL-KDD, ISCX & NIDS & 2020 & \cite{meira2020performance}\\
\hline
Unsupervised & IF, HBOS, CBLOF, K-Means 
 & BRO DNS, BRO CONN & NIDS & 2020 & \cite{hariharan2020camlpad}\\
\hline
Supervised & DNN & KDD-Cup99, NSL-KDD, UNSW-NB15 & NIDS & 2020 & \cite{choudhary2020analysis}\\
\hline
Supervised & KNN, RF, SVM-rbf, DNN, ResNet-50, one-vs-all classifier, multiclass classifier & NSL-KDD & NIDS & 2020 & \cite{wang2020explainable}\\
\hline
Supervised & NB, DT, RF, ANN & KDD-Cup99 & NIDS & 2020 & \cite{alqahtani2020cyber}\\
\hline
Supervised & SVM, MLP, NB, DT & ADFA-LD & HIDS & 2017 & \cite{subba2017host}\\
\hline
Supervised & CNN & NGIDS-DS, ADFA-LD & HIDS & 2017 & \cite{tran2017approach}\\
\hline
Semi-Supervised & SC4ID & ADFA-LD, UNM dataset & HIDS & 2018 & \cite{marteau2018sequence}\\
\hline
Supervised & GRU, LSTM, CNN+GRU & ADFA-LD & HIDS & 2019 & \cite{chawla2018host}\\
\hline
Supervised & LR, SVM, DT, RF, ANN & DS2OS traffic traces & HIDS & 2019 & \cite{evangelou2020anomaly}\\
\hline
Supervised & NN, DT, linear discriminate analysis with the bagging algorithm & NSL-KDD & HIDS & 2019 & \cite{besharati2019lr}\\
\hline
Supervised & NB, LR, KNN, SVM, IntruDTree & Kaggle cybersecurity datasets & HIDS & 2020 & \cite{sarker2020intrudtree}\\
\hline
\label{table:mlmodels-new}
\end{longtable}
\end{center}

As can be seen from the Table \ref{table:mlmodels-new}, most existing studies focus on NIDS. More datasets have been created for NIDS and different machine learning algorithms have been explored. However, compared to NIDS, HIDS is more of a challenge due to \cite{jose2018survey}: (1) More labeled data is required to reduce the false positive alarm rate. (2) It is difficult to design an efficient HIDS which can prevent outgoing denial-of-service attacks. (3) In a shared system environment, the HIDS needs to work as an independent module since the shared parameters may cause the attack. Nowadays, HIDS are becoming more important and play a major role in most of the intrusion detection systems \cite{jose2018survey}. Even though some studies have been conducted on HIDS \cite{subba2017host,tran2017approach,marteau2018sequence, chawla2018host, evangelou2020anomaly,sarker2020intrudtree}, more attention should be paid to HIDS. The applications of SOTA techniques, such as combining powerful language models with deep learning, might be a promising direction.  

\subsection{Datasets for intrusion detection}

As shown in Table \ref{table:mlmodels-new}, many datasets have been created for intrusion detection \cite{ring2019survey, khraisat2019survey}. Datasets for NIDS mainly include information from the packet itself and aim at detecting the malicious activity in network traffic using the content of individual packets, while datasets for HIDS usually include information about events or system calls/logs on a particular system with the purpose of detecting vulnerability exploits against a target application or computer system \cite{chawla2018host}. We conducted an investigation into the popular datasets used for NIDS and HIDS, as shown in Table \ref{table:idsdatasets}. 

\begin{center}
\begin{longtable}{p{2.5cm}p{3.5cm}p{3.5cm}p{1.2cm}p{1.2cm}p{1.2cm}p{0.8cm}}
\caption{\centering Overview of public datasets for IDS. The detailed description of the datasets are presented in our GitHub repository.} \\
\hline
\textbf{Dataset}& \textbf{Volume}& \textbf{Information}& \textbf{Format}& \textbf{Labeled}& \textbf{Balanced}& \textbf{Year}\\
\hline \hline
\multicolumn{7}{ c }{NIDS}\\
\hline
KDD-Cup99 \cite{stolfo2000cost} & 4.9 millions for training, 2 million for testing & 41 features, 20 types of attacks & packet, logs & yes & no & 1999 \\
\hline
Kyoto 2006 \cite{song2006description} & 3,054,682 for training, 1,563,923 for testing &  14 features derived from the KDD-Cup99 and 10 additional features & packet, logs & yes & yes & 2006\\
\hline
DARPA-2009 \cite{darpa2009} & 673,931 records for training and 74,880 records for testing & 16 network features and 26 packet features; 7000 pcap files & packets & yes & no & 2009\\
\hline
NSL-KDD \cite{tavallaee2009detailed} & 125,973 for training, 22,544 for testing  & 41 features, 22 types of attacks & packet, logs & yes & no & 2009 \\
\hline
ISCX \cite{shiravi2012toward} & 30,814 normal and 15,375 attack traces for training, 13,154 normal and 6,580 attack traces for testing & 1.5 million network traffic packets, with 20 features and covered seven days of network activity & packets & yes & no & 2012\\
\hline
UNSW-NB15 \cite{moustafa2015unsw} & 175,341 records for training, 82,332 records for testing & 49 features in pcap file format and 9 categories of attacks & packets & yes & no & 2015\\
\hline
NGIDS-DS \cite{haider2017generating} & 631,85,992 records for training and 34,987,493 records for testing & 88,791,734 records for benign and 1,262,426 records for malicious activities. 7 features for ground-truth cs; 9 features for the 99 csv files of host logs; 18 features for NGIDS.pcap & packet, logs & yes & no & 2016\\
\hline
CICIDS2017 \cite{sharafaldin2018toward}  & 75,561 records for training and 25,187 records for testing & 3,119,345 instances and 83 features containing 15 class labels & packets & yes & no & 2017\\
\hline \hline
\multicolumn{7}{ c }{HIDS}\\
\hline
DARPA 98/99 \cite{DARPA-dataset} & 4,898,431 records for training and 2,984,154 records for testing & 97,277 normal and 311,744 intrusion traces with 41 features and classes labeled as either normal or any of the 22 types of attacks & packet, logs & yes & no & 1998 \\
\hline
UNM dataset \cite{hofmeyr1998intrusion} & 627 system-call sequences for training and 3,136 system-call sequences for testing & 4,298 normal traces and 1,001 intrusion traces and 467 features  & logs & yes & no & 1998\\
\hline
KDD99 \cite{stolfo2000cost} & 494,021 for training and 311,029 for testing & 1,033,372 normal and 4,176,086 attack traces; 41 features & packets & yes & no & 1999 \\
\hline
NSL-KDD \cite{tavallaee2009detailed} & 1,152,281 distinct records from KDD99: 860,725 normal and 291,556 attack traces & 41 features per record & packets & yes & no & 2000\\
\hline
ADFA-LD \cite{creech2013semantic} & 833 traces and 308,077 system calls for training, 4,373 traces and 2,122,085 system calls for testing & 746 trace attack sequences and 317,388 system call attack sequences  & & yes & no & 2014\\
\hline
ADFA-WD \cite{creech2013semantic} & 355 traces and 13,504,419 system calls for training, 1,827 traces and 117,918,735 system calls for testing & 5,542 trace attack sequences and 74,202,804 system call attack sequences & & yes & no & 2014\\
\hline
DARPA-2009 \cite{darpa2009} & 673,931 records for training and 74,880 records for testing & 16 network features and 26 packet features; 7000 pcap files & packets & yes & no & 2009\\
\hline
ADFA-IDS \cite{vceponis2018towards} & 308,077 system calls and 833 traces for training, 212,2085 system calls and 4,372 traces for testing & 15 attack types and 36,636 malicious system call traces & logs & yes & no & 2013\\
\hline
NGIDS-DS \cite{haider2017generating} & 313,926 records with 7 attributes in ground-truth csv file;1,262,426 attack and 88,791,734 normal with 9 attributes; 1,094,231 capture packets with 18 unique IPs in NGIDS.pcap file & cyber normal and abnormal traffic scenarios for different enterprises & packet, host logs & yes & no & 2017\\
\hline
DS2OS traffic traces \cite{pahl2018all} & 61.52 MB & traces captured in the IoT environment DS2OS for different services & -- & yes & yes & 2018\\
\hline
Kaggle cybersecurity datasets \cite{sarker2020intrudtree} & 25,000 instances & 3 qualitative features and 38 quantitative features & -- & yes & yes & 2020\\
\hline
\label{table:idsdatasets}
\end{longtable}
\end{center}

Generally, the following properties are required when creating an IDS dataset: (1) Normal user behavior. The quality of an IDS is primarily determined by its attack detection rate and false alarm rate. Therefore, the presence of normal user behavior is indispensable for evaluating an IDS \cite{ring2019survey}. (2) Attack traffic. The attack types in different scenarios varies, so it is necessary to clarify the attacks in the IDS dataset. (3) Format. An IDS dataset can be in different formats such as packet-based, flow-based, host-based log files, etc. (4) Anonymity. Some of the information is anonymized due to privacy concerns, and this property indicates which attributes will be affected. (5) Duration. The recording time (e.g., daytime vs. night or weekday vs. weekend) of the dataset is indicated since a behavior might be regarded as an attack only when it occurs in a specific duration. (6) Labeled. Labeled datasets are necessary for training supervised learning and semi-supervised learning models and for evaluating supervised learning, semi-supervised learning, and unsupervised learning models. (7) Other information, such as attack scenarios, network structure, IP addresses, recording environment, download URL, are also useful. Quality issues can easily appear in the above information. Those data quality issues, if not checked and eliminated appropriately, will greatly affect the intrusion detection performance. However, few studies have discussed the qualities of IDS datasets \cite{shiravi2012toward, haider2017generating} for machine learning, although data quality issues, such as duplication and imbalance, have been reported in the KDD dataset \cite{tavallaee2009detailed}.   

\subsection{Data quality evaluation and assurance for machine learning}
Poor data quality has a direct impact on the performance of the machine learning system that is built on the data. For example, a face recognition-based gender classification system that was implemented with machine learning algorithms produced a 0.8\% error rate when recognizing the faces of lighter-skinned males, but went as high as a 34.7\% error rate when recognizing the faces of darker-skinned females \cite{pmlr-v81-buolamwini18a}. The problem was due to a significant imbalance of the training datasets in skin colors \cite{pmlr-v81-buolamwini18a}. Recently, a study showed that ten of the most commonly-used computer vision, natural language, and audio datasets had serious data quality issues \cite{northcutt2021pervasive}. The computing rule of ``garbage in, garbage out'' is also applicable to machine learning-based anomaly detection. However, there is not much research on the analysis of low quality training data and its impact on machine learning-based anomaly detection.

One of the fundamental issues might be the noisy labels. To investigate the impact of the quality of the labeling process on the performance of the machine learning-based network intrusion detection, Laur{\'i}a and Tayi compared two classification algorithms (DT and NB) which were trained on poor quality data  \cite{Laura2008StatisticalML}. The experiments showed that data with totally clean labels may not be required to train a classifier that performs at an acceptable level as a detector of network intrusions \cite{Laura2008StatisticalML}. Class imbalance is another common issue in IDS. As pointed out by Sahu et al., both of the datasets CIDDS and KDD are imbalanced in classes, and the distribution of KDD is even less uniform: the two most dominant classes both have more than 40,000 instances while the least dominant 16 classes have less than 1,000 instances \cite{sahu2020data}. Their experiments demonstrated that data balancing cannot improve its performance if the training dataset is less uniform and data balancing will benefit the neural network if improving the predictive accuracy of less dominant classes is desired \cite{sahu2020data}. Oversampling and undersampling have been used to deal with the class imbalance in IDS \cite{Divekar2018BenchmarkingDF}. Missing information, duplicate data, attack diversity, dataset difficulty, and feature sparsity can be other factors that reduce the performance of machine learning-based IDS \cite{Verma2019DataQF, sahu2020data}. Different dimensions have been proposed to measure the data quality for machine learning systems. Fan argued that data consistency, data de-duplication, information completeness, data currency, and data accuracy are central to data quality \cite{fan2015data}. These dimensions are related to the data itself. The dimensions related to users include accessibility, integrability, interpretability, rectifiability, relevance, and timeliness \cite{KARR2006137}. Recently, Chen, Chen, and Ding defined ``data quality'' as a measurement of data for fitting the purpose of building a machine learning system \cite{9282280}. Dimensions, such as comprehensiveness, correctness, and variety, are critical to evaluate the data quality for machine learning systems \cite{9282280}. Gradient boosted decision tree, data filtering, SVM, and transfer learning have all been investigated for data quality assurance and improvement \cite{wu2015robust, 9282280, saez2016influence}. 

\section{Data preparation and quality requirement for intrusion detection}
\label{sec:datapreparation}

As discussed above, data quality is crucial for machine learning-based IDS. However, each stage of the data preparation can be plagued with problems of data quality, such as scattered presence of the data source, insufficient data during data collecting, label noise during the data annotation, and overlapping issues during the training-testing data splitting. Therefore, it is necessary to identify the dataset attributes and potential, then develop a guideline for quality assurance during data preparation.  

\subsection{Data preparation workflow}

\begin{figure}[h]
    \centering
    \includegraphics[width=1.0\textwidth]{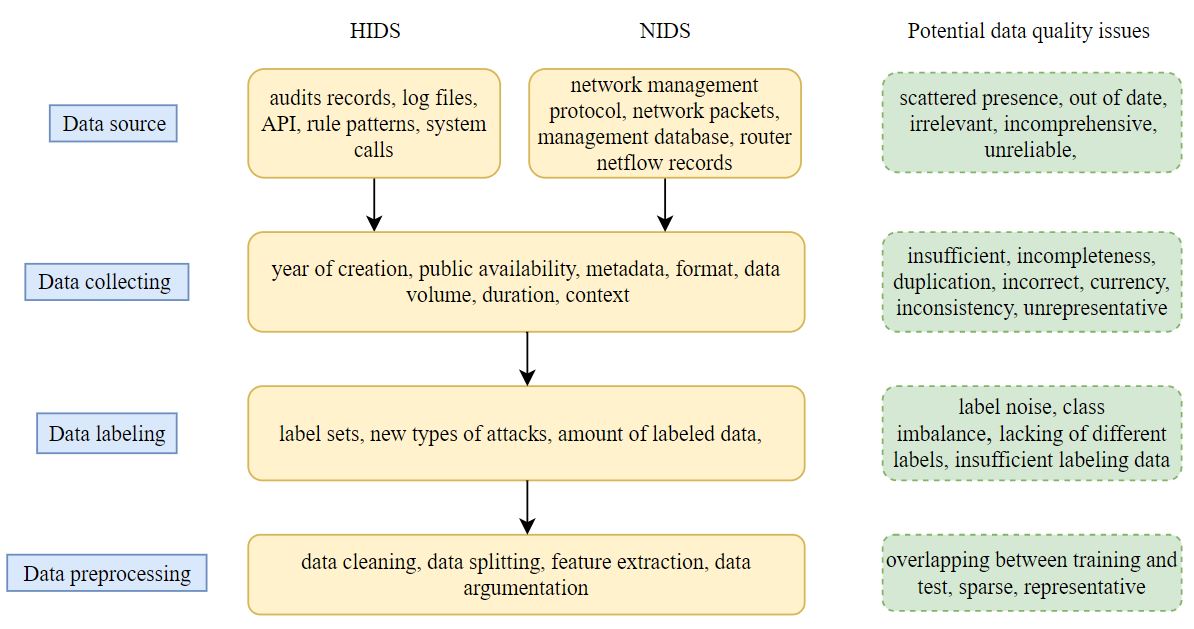}
    \caption{Data preparation workflow for machine learning-based IDS}
    \label{fig:1}
\end{figure}

Data preparation for machine learning-based IDS mainly includes four steps: (1) Selecting a data source or multiple data sources. (2) Collecting the data from the selected data source. (3) Labeling the data for training and testing. (4) Preprocessing the data as the model input. The workflow is shown in Figure \ref{fig:1}. The data sources for HIDS and NIDS are different: data for HIDS can be collected from audit records, log files, the application program interface (API), rule patterns, and system calls, while data for NIDS is usually collected from the simple network management protocol (SNMP), network packets (TCP/UDP/ICMP), management information base (MIB), and router NetFlow records. Data can be collected from one data source or by integrating multiple data sources. The data source is the foundation of accessing high quality data. Once the data source is confirmed, additional information should be collected, such as metadata, format, duration, etc., for further analysis. The data collecting procedure should be well designed to ensure the data quality. For example, when collecting sequential events, they should be correctly organized by their order. Data labeling is an essential step for supervised machine learning-based IDS. Most existing machine learning algorithms make the assumption that the training data feeding the algorithms is accurate (has no errors). However, errors in label data entry, lack of precision in expert judgment, and imbalanced data distribution in different categories during the process of labeling the training examples can impact the predictive accuracy of the classification algorithms \cite{Laura2008StatisticalML}. Data preprocessing aims at removing the outliers, cleaning the data, extracting the useful features, and splitting the data for training and test. This process, if handled inappropriately, will cause data quality issues, such as data sparsity and bias, and overlapping between training and test, which can also reduce the performance of the machine learning algorithms.         

\subsection{Dataset properties and attributes}

As pointed out by \cite{ring2019survey}, certain properties of a dataset should be collected and evaluated based on a specific scenario. We believe that to build a machine learning-based IDS, the following information should be collected during the data curation. 

\paragraph{General information.} General information of a dataset might include year of creation, public availability, metadata, format, and data volume. Both network traffic and system calls face the issue of concept drift over time, and new attacks might appear in any scenario. A machine learning model built in 1990 might not be useful to predict the data in 2021. Therefore, the year of creation (age) of an intrusion detection dataset is critical for deciding the scope of an IDS. Intrusion detection datasets should be publicly available to serve as a basis for comparing different intrusion detection methods and for quality evaluation by third parties \cite{ring2019survey}. Metadata, such as network structure, hosts, IP addresses, configuration of the clients, and attack scenarios, can provide users content-related explanations of the results. IDS datasets can be roughly divided into three formats: (1) packet, (2) flow, and (3) logs. The processing for different data formats varies. The format is directly associated with the volume of data. Data volume can be described by the number of packets, flows, points, and instances, which is crucial when selecting machine learning models. 

\paragraph{Duration.} Duration refers to the timestamps of when that data was collected. For example, the MIT dataset includes live normal data for lpr for two weeks using 77 hosts, while the ISCX dataset consists of the seven days of network activity (normal and malicious) from the Information Security Center of Excellence (ISCX) at the University of New Brunswick. As mentioned in \cite{macia2018ugr}, in order to enable the evaluation of detection algorithms that consider the cyclostationary evolution of traffic (i.e., differences in traffic between daytime/nighttime or weekdays/weekends), a long duration trace is needed.

\paragraph{Context.} Context is related to the recording environment. It delineates the network environment and conditions in which the datasets are captured. As for NIDS, the kind of traffic, type of network, and complete network indicate the context information. HIDS is generally a software component and is located on the system being monitored, so is typically monitoring a single system. The context information allows deeper understanding of processes and activities.    

\paragraph{Normal traces and types of attacks for intrusion traces.} This information is related to data labeling, which is the foundation of machine learning-based IDS. The training/validate/testing data should be correctly labeled as normal or not. In attack records, the type of attack is also needed \cite{macia2018ugr}. Maci{\'a}-Fern{\'a}ndez et al. applied the following strategy to label an instance: a) an attack label for the flows that they positively know correspond to an attack, b) a normal label for those that are generated synthetically with normal patterns, and c) a background label for those which no one knows exactly if they are attacks or not \cite{macia2018ugr}. Ring et al. summarized the specific types of attacks used in different datasets \cite{ring2019survey}. 

\paragraph{Features.} The input of machine learning-based IDS is a collection of data instances, each of which is represented by only one feature or a group of features. As shown in Table \ref{table:idsdatasets}, most of the datasets include the feature information, which can be qualitative features or quantitative features, network features or packet features, n-gram features, and other features. Features are the key components for developing an effective IDS.      

\subsection{Dataset quality principles and dimensions}

\label{sec:dataqualitydimension}

Different principles, metrics, and dimensions have been proposed to measure data quality \cite{wang1996beyond, batini2009methodologies, fan2015data, chen2019practical}. However, few of them are discussed in the context of building machine learning systems \cite{9282280}. As for machine learning-based IDS, central to data quality are reputation, relevance, comprehensiveness, timeliness, variety, accuracy, consistency, and de-duplication. We will discuss these dimensions in the following: 

\paragraph{Reputation.} ``Reputation'' is related to reliability, believability, and trustworthiness, which lays the foundation of the data quality. Reputation is evaluated using an information-theoretic concept, the Kullback-Leibler distance. Data for IDS can be collected from different sources, including host logs, network traffic, and application data. For a single data source, we can use a collective measure of trustworthiness (in the sense of reliability) based on the referrals or ratings from members in a community to evaluate the reputation \cite{josang2007survey}. PageRank is another method for reputation measurement \cite{bradai2013game}. If we generate the dataset by data fusion of multiple data sources, an “opinion” (a metric of the degree of belief) can be generated to represent the uncertainty in the aggregated result.  

\paragraph{Relevance.} ``Relevance'' indicates why the data is collected. Data should be collected and evaluated by ``fit for purpose'' \cite{9282280}. For example, HIDS data should be collected from the host system and audit sources, such as operating system, window server logs, firewall logs, application system audits, or database logs. While NIDS data should be extracted from a network through packet capture, NetFlow, and other network data sources. Moreover, if the IDS targets a specific type of attack, such as DDoS attacks, then the data also needs to be relevant to this attack. 

\paragraph{Comprehensiveness.} Existing machine learning-based intrusion detection systems frequently suffer from the issues of bias and lacking of robustness, which are mainly caused by the incomprehensiveness of the dataset. For example, the dataset does not contain balanced data for different normal or attack behavior, and the data cannot represent various features. Comprehensiveness requires a dataset to contain all representative samples from the population \cite{9282280}. For example, the NSL-KDD dataset includes a total of 39 attacks where each one of them is classified into one of the following four categories: DoS, R2L, U2R, and probe. Suppose that all the attacks should have an instance in the training set. However, 17 of these attacks are introduced only in the testing set. This dataset cannot then be considered as comprehensive. The importance of the comprehensiveness of data to machine learning, especially deep learning, is well understood since a deep learning model normally includes millions of parameters that needs a large amount of data to train it. If a comprehensive dataset with as many different types of attacks included, similar to the ImageNet for computer vision, can be developed for IDS, it would enhance the performance of machine learning-based IDS.  

\paragraph{Timeliness.} ``Timeliness'' (also called ``currency'') refers to the extent to which the age of the data \cite{wang1996beyond} is appropriate for the IDS task. Timeliness is an important factor to affect the performance of machine learning models since new types of attacks are emerging constantly, and some existing datasets, such as DARPA and KDD99, are too old to reflect these new attacks. Although the ADFA dataset contains many new attacks, it cannot be considered as comprehensive. For that reason, testing of machine learning models for IDS using DARPA, KDD99, and ADFA datasets does not offer a real evaluation and could result in inaccurate claims for their effectiveness \cite{khraisat2019survey}. Ideally, datasets should include most of the common attacks and correspond to current network environments \cite{liu2019machine}. 

\paragraph{Variety.} ``Variety'' concerns the coverage of the instances on the selected features. For example, the KDD-Cup99 dataset has 41 features, and it is supposed to be a normal distribution in the selected features with known mean and standard deviation in the real world. Otherwise, it will induce the data sparsity issue. Moreover, to improve the robustness in machine learning models, the instances in the validate data and test data should have enough variety to test the training model. Variety is considered as a subset of comprehensiveness in the scenario of constructing a machine learning system for intrusion detection. 

\paragraph{Accuracy.} According to the definition from Wang and Strong \cite{wang1996beyond}, ``accuracy'' means ``the extent to which data are correct, reliable and certified.'' Generally, accuracy can be distinguished by two aspects: syntactic and semantic \cite{liu2019machine}. However, for machine learning systems, the labeling accuracy should also be taken into consideration. Syntactic accuracy aims to check whether a value is any one of the values of, or how close it is to, the elements of the corresponding defined feature, while semantic accuracy requires an instance to be semantically represented appropriately. Labeling accuracy means that an instance should be correctly labeled as normal or as any type of attack. 

\paragraph{Consistency.} ``Data consistency'' refers to the validity and integrity of data representing real-world entities \cite{fan2015data}. It aims to detect errors, such as inconsistencies and conflicts in the data, typically identified as violations of data dependencies \cite{fan2015data}. For example, the system call for HIDS should be represented to ensure the sequential order, and the value of an attribute’s domain (feature) should be constrained to the range of admissible values. 

\paragraph{De-duplication.} Data de-duplication is the problem of identifying the same instances. It is a longstanding issue that has been studied for decades. Duplication was usually caused by ``data from a large number of (heterogeneous) data sources was not fused appropriately.'' Recently, Panigrahi and Borah reported that the CICIDS2017 dataset contains many redundant records which seems to be irrelevant for training any IDS \cite{panigrahi2018detailed}. However, duplication is not necessarily a data quality issue for machine learning-based IDS. For example, if one of the records (a data instance and its label) appears multiple times in training data, it reflects the probability of this attack behavior. This will not be considered as a duplication issue. However, having a large overlap between the training and the test data can potentially introduce bias in the model and contribute to high inaccuracy, as pointed out by Chen et al. \cite{9282280}. Transfer learning \cite{9282280}, the distance-based approach \cite{fan2015data}, the rule-based approach \cite{fan2015data}, and probabilistic \cite{fan2015data} can be used for data de-duplication. 

In Section \ref{sec:4}, we will conduct a case study of the datasets used in the experimental study using the data quality dimensions discussed above. Based on the case study, we will discover how data quality affects the performance of machine learning-based IDS, thereby understanding the strategies to assure data quality. 

\section{Experiment on machine learning-based intrusion detection}
\label{sec:4}
\subsection{Data collection}

\subsubsection{Data cleaning and prepossessing}

\label{subsubsection:preprocessing}

Since the goal of this case study is to develop a host-based intrusion detection system (HIDS), we conduct machine learning experiments on the UNM, MIT, and ADFA-LD datasets. A detailed introduction of these datasets can be found in our GitHub repository. Regardless of the slight difference in ADFA and UNM data formats, we use similar data cleaning and augmentation techniques to create a dataset for each class. Processing data for pre-trained language models vectors, such as BERT \cite{devlin2019bert} and GPT-2 \cite{radford2019language}, is similar to normal machine learning algorithms. We use tokenizer to parse data into system call sequences with a length of six. Since the UNM dataset contains system calls from concurrently running processes, we group them by PID to ensure their sequential order. On the other hand, as the ADFA-LD dataset is already organized by different processes, and there is no PID provided, we do not need to group system calls together. Once the data is in order, we tokenize them into a sequence of six. By tokenizing into a sequence of 6-grams, we increase the amount of data for training as well as testing purposes. In addition, the number of features will decrease when a trace is tokenized into smaller chunks, and this will increase the efficiency of training as well as testing performances \cite{hofmeyr1998intrusion}.








 
We clean the data by removing rows or sequences that appear in both normal and intrusion data. This step draws distinctive characteristics between the two classes and effectively boosts the machine learning performance. A row with normal sequence is labeled 0, whereas the one with intrusion sequence is labeled 1. We use normal data and intrusion data from each dataset to create a sample pool. If it is imbalanced, we use the bootstrapping method to create a balanced sample of normal sequences and intrusion sequences. Then, we split the sample into training and testing sets in a 70-30 ratio. By training with only signature sequences from both classes, we increase the model accuracy and recall (true positive rate, TPR) as well as decrease its false positive rate (FPR). 

\subsubsection{Descriptive analysis of the datasets}

\begin{figure}[htbp]
\centering
\begin{subfigure}[]{0.95\textwidth}
    \includegraphics[width=1.0\textwidth]{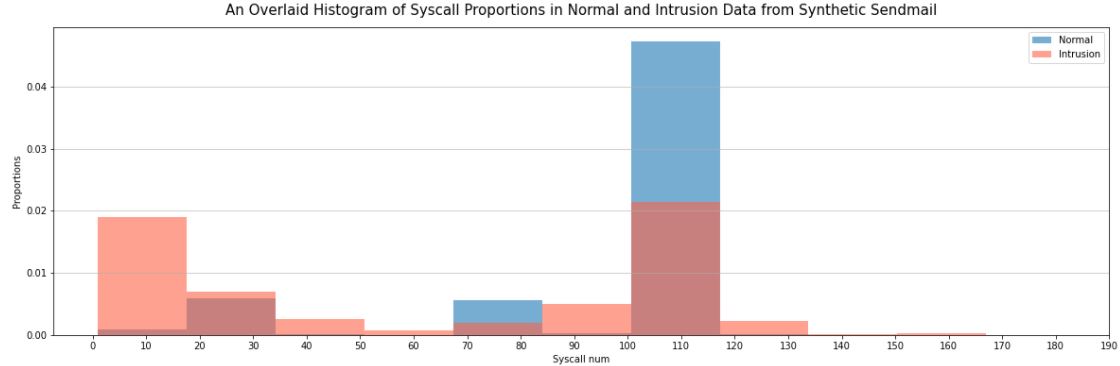}
    \caption{The Original UNM Synthetic Sendmail Dataset}
\label{fig:figure2_1}
\end{subfigure}\hfill

\begin{subfigure}[]{0.95\textwidth}
    \includegraphics[width=1.0\textwidth]{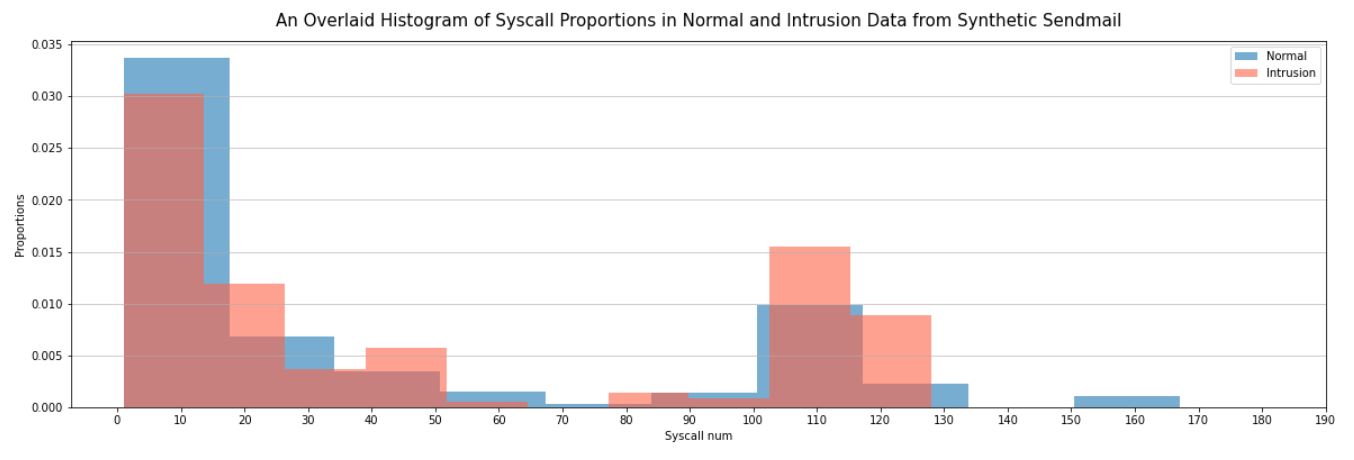}
    \caption{The Processed UNM Synthetic Sendmail Dataset}
\label{fig:figure3_1}
\end{subfigure}\hfill

\captionsetup{justification=centering,margin=0.5cm}
\caption{Overlaid Histograms of the Original and the Processed System Calls in Normal and Intrusion Data of the UNM Synthetic Sendmail Dataset.}
\label{figure2_99}
\end{figure}

In order to compare the difference between the two classes, we graph an overlaid histogram of normal data and intrusion data from each dataset in Figure \ref{figure2_99}. We use all of the UNM (except Sendmail), MIT Live Lpr, and ADFA-LD datasets. Each dataset has two histogram versions. However, we only include the histograms of the processed and unprocessed UNM Synthetic Sendmail dataset. The histograms of the rest of the datasets along with their descriptive analyses are included in our GitHub repository. The first histogram version displays the dispersion of original traces that have not been cleaned nor processed yet, so it can show the actual difference between normal and intrusion sequences. After cleaning duplicated sequences that exist in both classes, the overlaid histograms of most datasets have changed significantly. Therefore, the second version exhibits a more distinguished difference between the two classes than the original data. The goal is to differentiate normal sequences from intrusion sequences as much as possible so that the algorithms can learn to distinguish them from one another. Therefore, the more distinctive the two classes are, the better the candidate algorithms can perform. 

Figure \ref{fig:figure2_1} shows that there is a slight difference between normal data and intrusion data from the Synthetic Sendmail dataset. However, most of them overlap each other. After cleaning duplicated sequences, the dispersion of normal system calls and intrusion system calls have changed in Figure \ref{fig:figure3_1}. Normal system calls have expanded from two wide ranges (68 to 83 and 101 to 118) to multiple specific ranges. Particularly, sequences that contain system call numbers ranging between 13 and 15, 64 and 77, 100 and 102, 128 and 134, and 150 and 177 are most likely normal. This increases the homogeneity of normal data and reduces false negatives. Additionally, after being processed, intrusion system calls have expanded to more specific ranges. For example, sequences that contain system call numbers from 17 to 27, 77 to 84, and 102 to 128 are most likely intrusion. This narrows down possible intrusion likelihoods and reduces false positives in overlapped system calls.

\subsection{Machine learning algorithms}

\paragraph{K-means.} K-means is a common algorithm for anomaly detection, which groups similar characteristics into a cluster. We choose the number of clusters to be two because there are two categories in this dataset: normal (0) and intrusion (1). The rest of the parameters of K-means are set as default from the scikit-learn library. K-means is trained with 9,800 sequences and tested with 4,200 sequences (roughly 7,000 normal and 7,000 intrusion sequences are split into training and testing sets; the rest of the models use the same data for training and testing) from both classes in each dataset.



\paragraph{Logistic Regression.} Logistic regression (LR) is another potential candidate for detecting anomaly using maximum likelihood. Since this is a supervised model, we use sequences from the training set with labels to train the model. The model is evaluated on the testing set, where the true labels are compared against the predicted ones to measure the model performance. All parameters of this model are set as default in the scikit-learn library. 


\paragraph{Support Vector Machine.} Support vector machine (SVM) is a supervised machine learning algorithm which detects anomaly by separating normal behavior from the other by using different kernel types (linear, polynomial, and RBF). We choose a polynomial kernel with 3 degrees and 1.0 regularization to represent a SVM model. 


\paragraph{Neural Network.} Neural network (NN) is efficient at learning underlying complex relationships because of its composite architecture. The model is comprised of three layers: an input layer of six features, where each of them represents a system call from a sequence of six; a ReLU layer with six hidden nodes containing extracted information from each system call; and a sigmoid output layer with two output nodes, where each one represents a probability of a sequence belonging to either class. Whichever class has a higher chance in the output node will be the predicted label of the given input sequence. 


\paragraph{Decision Tree.} Decision tree (DT) relies on a set of rules, derived from the training process, to partition data into groups that are as homogeneous as possible. The goal is to generate a DT to classify normal sequences from intrusion sequences at the lowest possible error rate. That way, it can be generalized on the testing set as well as future unseen data. For every split, the model considers, at most, a square root number of features and decides where to split using GINI criterion with a minimum of ten samples per split and five samples per leaf. Each leaf node contains sequences, where most of them belong to the same class. 


\paragraph{Random Forest.} Random forest (RF) distinguishes normal sequences from intrusion sequences using the ensemble learning method. RF creates multiple DTs to classify the same observation. The final decision is based on the wisdom of the crowd that is the majority predicted class of that particular observation. Similarly, we set the same parameters as the DT model above. The only difference is that the model allows bootstrap samples to build multiple trees. 


\paragraph{K Nearest Neighbor.} K nearest neighbor (KNN) classifies an observation by counting the majority classes of the nearest neighbors or sequences. This model enhances its performance by minimizing inter-variability within a group while maximizing inter-variability between different groups. For simplicity, we choose the number of neighbors to be three. Still, any odd number of k is recommended to avoid a tie situation. Additionally, the weights parameter is set to be uniform, where all observations in each neighborhood are weighted equally regardless of their distances. 


\paragraph{Na\"ive Bayes.} Na\"ive Bayes (NB) is known for classifying data based on conditional probabilities without making any assumptions. Given lots of labeled sequences, we hope that this model can efficiently distinguish intrusion sequences from normal sequences. A predicted label is determined based on the highest probability of a class. The parameters of this model are set as default from the scikit-learn library. 


\paragraph{BERT.} Bidirectional encoder representations from transformers (BERT) was developed by the Google AI language, using transformer architecture with attention mechanisms for learning context. BERT is a bidirectional unsupervised language model, which takes the texts before and after the token into contextual account. There are two versions of BERT models based on structure size: BERT Base with 110 million parameters and BERT Large with 345 million parameters \cite{devlin2019bert}. In this research, the processed string in Section \ref{subsubsection:preprocessing} is passed to a tokenizer and each string of system call sequence is tokenized by the BERT tokenizer (bert-base-uncased). Each sequence is padded with [CLS] and [SEP] at the beginning and the end of the sequence. The output of the BERT model is then passed to a linear network with 768 neurons and the output of the linear layer is passed to a single neuron with SoftMax activation. The model is trained and tuned, using normal and intrusion sequences from each dataset, with 16 epochs and 16 batch size. 

\paragraph{GPT-2.} Generative pretrained transformers (GPT) was developed by Microsoft's Open AI, using the decoder part of the transformer architecture \cite{radford2019language}. There are four types of GPT-2 depending on the model structure size, which can be small – 117M parameters, medium – 345M parameters, large – 762M parameters, and extra large – 1542M parameters. GPT-2 is auto-regressive which reuses its own output as an input sequence. We use small GPT-2 to train on processed data from both classes of each dataset using 16 epochs and 16 batch size. Unlike BERT, GPT-2 does not require an additional output layer; instead, it outputs the sequence's likelihood in each class.

\subsection{Experimental results}

\subsubsection{Overall results}

Table \ref{table:overall-results} shows the performance of different machine learning algorithms on 11 datasets from UNM, MIT, and ADFA-LD regarding performance indicator accuracy, recall, precision, macro-f1 score, false positive rate (meaning the percentage of malicious behaviors that are labeled as benign behaviors, FPR), and AUC score \cite{khraisat2019survey}. Our goal is to find the best candidate model with high accuracy and high recall but low FPR. High performance measures are in bold so that we can easily identify the best candidate model. We make the following observations:

\begin{itemize}
    \item Almost all of the algorithms achieved a better performance on the Synthetic Lpr, Live Lpr, Xlock, Live Named, Inetd, and Stide datasets than the others, indicating that the data quality of the former group of datasets might be higher than the latter group.   
    \item Decision tree, Random forest, KNN, BERT, and GPT-2 are the best candidate algorithms since they achieved higher accuracy, recall, and precision, yet at a lower false positive rate on all datasets.
    \item BERT and GPT-2 are the best algorithms for HIDS on all of the datasets since they outperformed other algorithms on almost all the metrics, and all the recalls are above 0.90 while the FPRs are below 0.06. 
    \item Class imbalance is not a major issue for the HIDS datasets used in this paper since we use the bootstrapping technique to generate a balanced sample of both classes. This sample is then split into training and testing sets with a 70-30 ratio to train and test the models. 
\end{itemize}

\begin{centering}
\begin{longtable}{l|l|l|l|l|l|l|l}
\caption{\centering Model performance regarding accuracy, recall, precision, macro-F1, FPR, and AUC score on different HIDS datasets.} \label{table:overall-results} \\
\hline
Dataset & Model & Accuracy & Recall & Precision & Macro-F1 & FPR & AUC score \\
\hline
\endfirsthead

\endlastfoot

\multirow{10}{4em}{Synthetic Sendmail} & K-means & 0.63 & 0.61 & 0.64 & 0.63 & 0.36 & 0.63 \\
& Logistic Regression & 0.63 & 0.57 & 0.62 & 0.61 & 0.35 & 0.63 \\ 
& SVM & 0.73 & 0.57 & 0.85 & 0.73 & 0.10 & 0.73\\ 
& Neural Network & 0.66 & 0.63 & 0.68 & 0.67 & 0.29 & 0.66 \\
& Decision Tree & \textbf{0.98} & \textbf{1.00} & \textbf{0.97} & \textbf{0.98} & \textbf{0.03} & \textbf{0.98} \\
& Random Forest & \textbf{0.99} & \textbf{1.00} & \textbf{0.98} & \textbf{0.99} & \textbf{0.02} & \textbf{0.99} \\
& KNN & \textbf{1.00} & \textbf{1.00} & \textbf{1.00} & \textbf{1.00} & \textbf{0.00} & \textbf{1.00}\\ 
& Na\"ive Bayes & 0.63 & 0.57 & 0.61 & 0.61 & 0.36 & 0.63 \\
& BERT & \textbf{1.00} & \textbf{1.00} & \textbf{1.00} & \textbf{1.00} & \textbf{0.00} & \textbf{1.00}\\ 
& GPT-2 Network & \textbf{1.00} & \textbf{1.00} & \textbf{0.99} & \textbf{1.00} & \textbf{0.01} & \textbf{1.00}\\ 
\hline

\multirow{10}{4em}{Synthetic Ftp} & K-means & 0.20 & 0.07 & 0.10 & 0.20 & 0.66 & 0.20 \\ 
& Logistic Regression & 0.74 & 0.77 & 0.74 & 0.74 & 0.28 & 0.74 \\ 
& SVM & 0.79 & 0.67 & \textbf{0.90} & 0.79 & \textbf{0.08} & 0.79\\ 
& Neural Network & 0.80 & \textbf{0.89} & 0.76 & 0.80 & 0.30 & 0.80\\
& Decision Tree & \textbf{0.99} & \textbf{1.00} & \textbf{0.99} & \textbf{0.99} & \textbf{0.02} & \textbf{0.99} \\
& Random Forest & \textbf{0.99} & \textbf{1.00} & \textbf{0.99} & \textbf{0.99} & \textbf{0.01} & \textbf{0.99} \\
& KNN & \textbf{0.99} & \textbf{1.00} & \textbf{0.99} & \textbf{0.99} & \textbf{0.01} & \textbf{0.99}\\
& Na\"ive Bayes & 0.77 & 0.82 & 0.75 & 0.77 & 0.28 & 0.77 \\
& BERT & \textbf{1.00} & \textbf{1.00} & \textbf{1.00} & \textbf{1.00} & \textbf{0.00} & \textbf{1.00}\\ 
& GPT-2 Network & \textbf{0.99} & \textbf{0.99} & \textbf{1.00} & \textbf{0.99} & \textbf{0.00} & \textbf{0.99}\\ 
\hline

\multirow{10}{4em}{Synthetic Lpr} & K-means & 0.55 & 0.10 & \textbf{0.93} & 0.54 & \textbf{0.01} & 0.54\\
& Logistic Regression & \textbf{0.97} & \textbf{0.99} & \textbf{0.95} & \textbf{0.97} & \textbf{0.05} & \textbf{0.97} \\ 
& SVM & \textbf{0.99} & \textbf{0.99} & \textbf{0.99} & \textbf{0.99} & \textbf{0.01} & \textbf{0.99} \\
& Neural Network & \textbf{0.97} & \textbf{0.99} & \textbf{0.95} & \textbf{0.97} & \textbf{0.05} & \textbf{0.97}\\
& Decision Tree & \textbf{0.99} & \textbf{1.00} & \textbf{0.98} & \textbf{0.99} & \textbf{0.02} & \textbf{0.99} \\
& Random Forest & \textbf{1.00} & \textbf{0.99} & \textbf{1.00} & \textbf{1.00} & \textbf{0.00} & \textbf{1.00} \\
& KNN & \textbf{1.00} & \textbf{1.00} & \textbf{1.00} & \textbf{1.00} & \textbf{0.00 }& \textbf{1.00} \\
& Na\"ive Bayes & \textbf{0.96} & \textbf{1.00} & \textbf{0.93} & \textbf{0.96} & \textbf{0.07} & \textbf{0.96} \\
& BERT & \textbf{1.00} & \textbf{1.00} & \textbf{0.99} & \textbf{1.00} & \textbf{0.01} & \textbf{1.00}\\ 
& GPT-2 Network & \textbf{1.00} & \textbf{1.00} & \textbf{1.00} & \textbf{1.00} & \textbf{0.00} & \textbf{1.00}\\ 
\hline

\multirow{10}{4em}{Live Lpr} & K-means & 0.86 & 0.76 & \textbf{0.96} & 0.86 & \textbf{0.03} & 0.86\\
& Logistic Regression & \textbf{0.98} & \textbf{1.00} & \textbf{0.97} & \textbf{0.98} & \textbf{0.03} & \textbf{0.98} \\ 
& SVM & \textbf{1.00} & \textbf{1.00} & \textbf{1.00} & \textbf{1.00} & \textbf{0.00} & \textbf{1.00}\\ 
& Neural Network & \textbf{0.98} & \textbf{1.00} & \textbf{0.95} & \textbf{0.98} & \textbf{0.05} & \textbf{0.98}\\
& Decision Tree & \textbf{1.00} & \textbf{1.00} & \textbf{1.00} & \textbf{1.00} & \textbf{0.00} & \textbf{1.00} \\
& Random Forest & \textbf{1.00} & \textbf{1.00} & \textbf{1.00} & \textbf{1.00} & \textbf{0.00} & \textbf{1.00} \\
& KNN & \textbf{1.00} & \textbf{1.00} & \textbf{1.00} & \textbf{1.00} & \textbf{0.00} & \textbf{1.00} \\
& Na\"ive Bayes & \textbf{0.96} & \textbf{1.00} & \textbf{0.93} & \textbf{0.96} & \textbf{0.07} & \textbf{0.96} \\
& BERT & \textbf{1.00} & \textbf{1.00} & \textbf{1.00} & \textbf{1.00} & \textbf{0.00} & \textbf{1.00} \\
& GPT-2 Network & \textbf{1.00} & \textbf{1.00} & \textbf{1.00} & \textbf{1.00} & \textbf{0.00} & \textbf{1.00} \\
\hline

\multirow{10}{4em}{MIT Live Lpr} & K-means & 0.70 & 0.54 & 0.79 & 0.70 & 0.15 & 0.70 \\ 
& Logistic Regression & 0.74 & 0.67 & 0.78 & 0.74 & 0.18 & 0.74 \\ 
& SVM & 0.76 & 0.69 & 0.79 & 0.75 & 0.17 & 0.75\\ 
& Neural Network & 0.75 & 0.70 & 0.77 & 0.75 & 0.20 & 0.75\\
& Decision Tree & \textbf{1.00} & \textbf{1.00} & \textbf{0.99} & \textbf{1.00} & \textbf{0.01} & \textbf{1.00}\\
& Random Forest & \textbf{1.00} & \textbf{1.00} & \textbf{1.00} & \textbf{1.00} & \textbf{0.00} & \textbf{1.00}\\
& KNN & \textbf{1.00} & \textbf{1.00} & \textbf{1.00} & \textbf{1.00} & \textbf{0.00} & \textbf{1.00} \\
& Na\"ive Bayes & 0.73 & 0.66 & 0.78 & 0.73 & 0.19 & 0.73 \\
& BERT & \textbf{1.00} & \textbf{1.00} & \textbf{1.00} & \textbf{1.00} & \textbf{0.00} & \textbf{1.00}\\
& GPT-2 Network & \textbf{1.00} & \textbf{1.00} & \textbf{1.00} & \textbf{1.00} & \textbf{0.00} & \textbf{1.00}\\
\hline

\multirow{10}{4em}{Xlock} & K-means & 0.37 & 0.31 & 0.36 & 0.37 & 0.56 & 0.37 \\ 
& Logistic Regression & 0.77 & 0.67 & 0.84 & 0.77 & 0.13 & 0.77 \\ 
& SVM & 0.84 & \textbf{0.97} & 0.77 & 0.84 & 0.29 & 0.84\\ 
& Neural Network & 0.79 & 0.74 & 0.82 & 0.79 & 0.17 & 0.79\\
& Decision Tree & \textbf{0.99} & \textbf{1.00} & \textbf{0.97} & \textbf{0.99} & \textbf{0.03} & \textbf{0.99} \\
& Random Forest & \textbf{0.99} & \textbf{1.00} & \textbf{0.98} & \textbf{0.99} & \textbf{0.02} & \textbf{0.99} \\
& KNN & \textbf{0.99} & \textbf{1.00} & \textbf{0.99} & \textbf{0.99} & \textbf{0.01} & \textbf{0.99} \\
& Na\"ive Bayes & 0.68 & 0.69 & 0.68 & 0.68 & 0.32 & 0.68 \\
& BERT & \textbf{1.00} & \textbf{1.00} & \textbf{1.00} & \textbf{1.00} & \textbf{0.00} & \textbf{1.00}\\ 
& GPT-2 Network & \textbf{0.99} & \textbf{1.00} & \textbf{0.99} & \textbf{0.99} & \textbf{0.01} & \textbf{0.99}\\
\hline

\multirow{10}{4em}{Live Named} & K-means & 0.21 & 0.38 & 0.29 & 0.21 & 0.96 & 0.21\\ 
& Logistic Regression & 0.83 & 0.75 & \textbf{0.90} & 0.83 & \textbf{0.08} & 0.83 \\ 
& SVM & \textbf{0.92} & \textbf{0.87} & \textbf{0.99} & \textbf{0.92} & \textbf{0.01} & \textbf{0.92}\\ 
& Neural Network & 0.84 & 0.72 & \textbf{0.95} & 0.84 & \textbf{0.04} & 0.84 \\
& Decision Tree & \textbf{1.00} & \textbf{1.00} & \textbf{1.00} & \textbf{1.00} &\textbf{0.00} & \textbf{1.00} \\ 
& Random Forest & \textbf{1.00} & \textbf{1.00} & \textbf{1.00} & \textbf{1.00} & \textbf{0.00} & \textbf{1.00} \\
& KNN & \textbf{1.00} & \textbf{1.00}  &\textbf{1.00} & \textbf{1.00} & \textbf{0.00} & \textbf{1.00} \\
& Na\"ive Bayes & 0.87 & 0.81 & \textbf{0.92} & \textbf{0.87} & \textbf{0.07} & \textbf{0.87} \\
& BERT & \textbf{1.00} & \textbf{1.00} & \textbf{1.00} & \textbf{1.00} &\textbf{0.00} & \textbf{1.00} \\ 
& GPT-2 Network & \textbf{1.00} & \textbf{1.00} & \textbf{1.00} & \textbf{1.00} &\textbf{0.00} & \textbf{1.00} \\ 
\hline

\multirow{10}{4em}{Login and Ps} & K-means & 0.52 & 0.64 & 0.53 & 0.51 & 0.61 & 0.51 \\ 
& Logistic Regression & 0.68 & 0.55 & 0.77 & 0.68 & 0.18 & 0.68 \\ 
& SVM & \textbf{0.92} & \textbf{0.94} & \textbf{0.92} & \textbf{0.92} &\textbf{0.09} & \textbf{0.92}\\ 
& Neural Network & 0.71 & 0.60 & 0.78 & 0.71 & 0.18 & 0.71\\
& Decision Tree & \textbf{1.00} & \textbf{1.00} & \textbf{1.00} &\textbf{1.00} & \textbf{0.00} & \textbf{1.00}\\
& Random Forest & \textbf{1.00} & \textbf{1.00} & \textbf{1.00} &\textbf{1.00} & \textbf{0.00} & \textbf{1.00}\\
& KNN &  \textbf{1.00} & \textbf{1.00} & \textbf{1.00} &\textbf{1.00} & \textbf{0.00} & \textbf{1.00} \\
& Na\"ive Bayes & 0.63 & 0.55 & 0.67 & 0.63 & 0.29 & 0.63 \\
& BERT & \textbf{1.00} & \textbf{1.00} & \textbf{1.00} &\textbf{1.00} & \textbf{0.00} & \textbf{1.00}\\
& GPT-2 Network & \textbf{1.00} & \textbf{1.00} & \textbf{1.00} &\textbf{1.00} & \textbf{0.00} & \textbf{1.00}\\
\hline

\multirow{10}{4em}{Inetd} & K-means & 0.68 & 0.82 & 0.64 & 0.68 & 0.46 & 0.68 \\ 
& Logistic Regression & 0.70 & 0.75 & 0.69 & 0.70 & 0.33 & 0.70 \\ 
& SVM & \textbf{0.96} &\textbf{0.95} & \textbf{0.97} & \textbf{0.96} & \textbf{0.03} & \textbf{0.96} \\
& Neural Network & 0.86 & \textbf{0.95} & 0.80 & 0.86 & 0.24 & 0.86\\
& Decision Tree & \textbf{1.00} &\textbf{1.00} & \textbf{1.00} & \textbf{1.00} & \textbf{0.00} & \textbf{1.00}\\
& Random Forest & \textbf{1.00} &\textbf{1.00} & \textbf{1.00} & \textbf{1.00} & \textbf{0.00} & \textbf{1.00}\\
& KNN & \textbf{1.00} &\textbf{1.00} & \textbf{1.00} & \textbf{1.00} & \textbf{0.00} & \textbf{1.00} \\
& Na\"ive Bayes & 0.71 & 0.81 & 0.68 & 0.71 & 0.39 & 0.71 \\
& BERT & \textbf{1.00} &\textbf{1.00} & \textbf{1.00} & \textbf{1.00} & \textbf{0.00} & \textbf{1.00}\\
& GPT-2 Network & \textbf{1.00} &\textbf{1.00} & \textbf{1.00} & \textbf{1.00} & \textbf{0.00} & \textbf{1.00}\\
\hline

\multirow{10}{4em}{Stide} & K-means & 0.54 & 0.66 & 0.53 & 0.54 & 0.59 & 0.54 \\
& Logistic Regression & 0.80 & 0.76 & 0.82 & 0.80 & 0.17 & 0.80 \\ 
& SVM & \textbf{0.91} &\textbf{0.99} & \textbf{0.85} & \textbf{0.91} & 0.18 & \textbf{0.91} \\
& Neural Network & 0.80 & 0.77 & 0.82 & 0.80 & 0.17 & 0.80\\
& Decision Tree & \textbf{1.00} &\textbf{1.00} & \textbf{1.00} & \textbf{1.00} & \textbf{0.00} & \textbf{1.00} \\
& Random Forest & \textbf{1.00} &\textbf{1.00} & \textbf{1.00} & \textbf{1.00} & \textbf{0.00} & \textbf{1.00} \\
& KNN & \textbf{1.00} &\textbf{1.00} & \textbf{1.00} & \textbf{1.00} & \textbf{0.00} & \textbf{1.00} \\
& Na\"ive Bayes & 0.76 & 0.80 & 0.74 & 0.76 & 0.29 & 0.76 \\
& BERT & \textbf{1.00} &\textbf{1.00} & \textbf{1.00} & \textbf{1.00} & \textbf{0.00} & \textbf{1.00} \\
& GPT-2 Network & \textbf{1.00} &\textbf{1.00} & \textbf{1.00} & \textbf{1.00} & \textbf{0.00} & \textbf{1.00} \\
\hline

\multirow{10}{4em}{ADFA-LD} & K-means & 0.61 & 0.62 & 0.60 & 0.61 & 0.39 & 0.61 \\
& Logistic Regression & 0.61 & 0.62 & 0.59 & 0.61 & 0.41 & 0.61 \\ 
& SVM & 0.65 & 0.53 & 0.69 & 0.65 & 0.23 & 0.65 \\
& Neural Network & 0.60 & 0.70 & 0.58 & 0.60 & 0.49 & 0.60 \\
& Decision Tree & \textbf{0.85} & \textbf{0.84} & \textbf{0.85} & \textbf{0.85} & \textbf{0.14} & \textbf{0.85} \\
& Random Forest & \textbf{0.88} & \textbf{0.87} & \textbf{0.89} & \textbf{0.88} & \textbf{0.11} & \textbf{0.88} \\
& KNN & \textbf{0.80} & \textbf{0.70} & \textbf{0.88} & \textbf{0.80} & \textbf{0.09} & \textbf{0.80}\\
& Na\"ive Bayes & 0.61 & 0.61 & 0.60 & 0.61 & 0.38 & 0.61 \\
& BERT & \textbf{0.94} &\textbf{0.92} & \textbf{0.95} & \textbf{0.94} & \textbf{0.05} & \textbf{0.94} \\
& GPT-2 Network & \textbf{0.93} &\textbf{0.92} & \textbf{0.94} & \textbf{0.93} & \textbf{0.06} & \textbf{0.93} \\
\hline
\end{longtable}
\end{centering}

\subsubsection{ROC Curve}
\begin{figure}[htbp]
\centering
\includegraphics[width=0.7\linewidth]{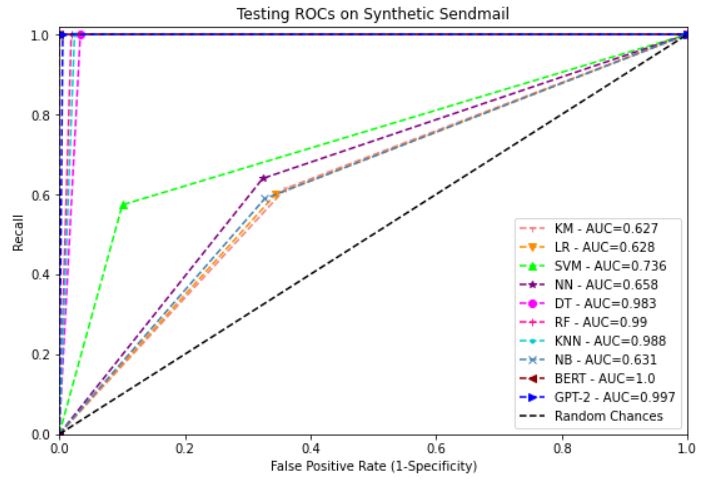}
\caption{ROC Curves on the Synthetic Sendmail dataset regarding different machine learning algorithms.}
\label{fig:roc-curve}
\end{figure}

Figure \ref{fig:roc-curve} shows the testing ROC curves of all candidate algorithms on the Synthetic Sendmail dataset. The ROC curves of the other datasets are included in our GitHub repository. Overall, RF, DT, KNN, BERT, and GPT-2 generate the highest results and outperform the others on all datasets. Their average AUC scores are 0.986 (RF), 0.982 (DT), and 0.980 (KNN), with 0.995 (BERT) and 0.993 (GPT-2) demonstrating near perfect performances. Moreover, their FPR are below 0.04 on all datasets except ADFA-LD. These five algorithms can truly learn and effectively distinguish between normal sequences and intrusion sequences. Contrarily, most of the other candidates only perform well on certain datasets, such as the Synthetic Ftp, Synthetic Lpr, Live Lpr, Xlock, Live Named, Inetd, and Stide datasets. Besides the five best candidates, there is no other algorithm that performs well when trained and tested on the Synthetic Sendmail, MIT Live Lpr, and ADFA-LD datasets. 

When training and testing the machine learning algorithms with the Synthetic Sendmail dataset, RF, DT, KNN, BERT, and GPT-2 surpass the other algorithms with AUC above 0.98. On the other hand, the other algorithms can only achieve AUC up to 0.74. The difference in performance has separated the candidate algorithms into two groups. One contains more effective algorithms, such as BERT, GPT-2, RF, DT, and KNN, and the other one contains average algorithms, such as SVM, K-means, LR. Next, we will use the log ratio of recall over FPR to determine the best and worst intrusion detection algorithms in Section \ref{sec:4.3.3}.

\subsubsection{Ratio of Recall over False Positive Rate}

\label{sec:4.3.3}
\begin{figure}[htbp]
\centering
\includegraphics[width=0.7\linewidth]{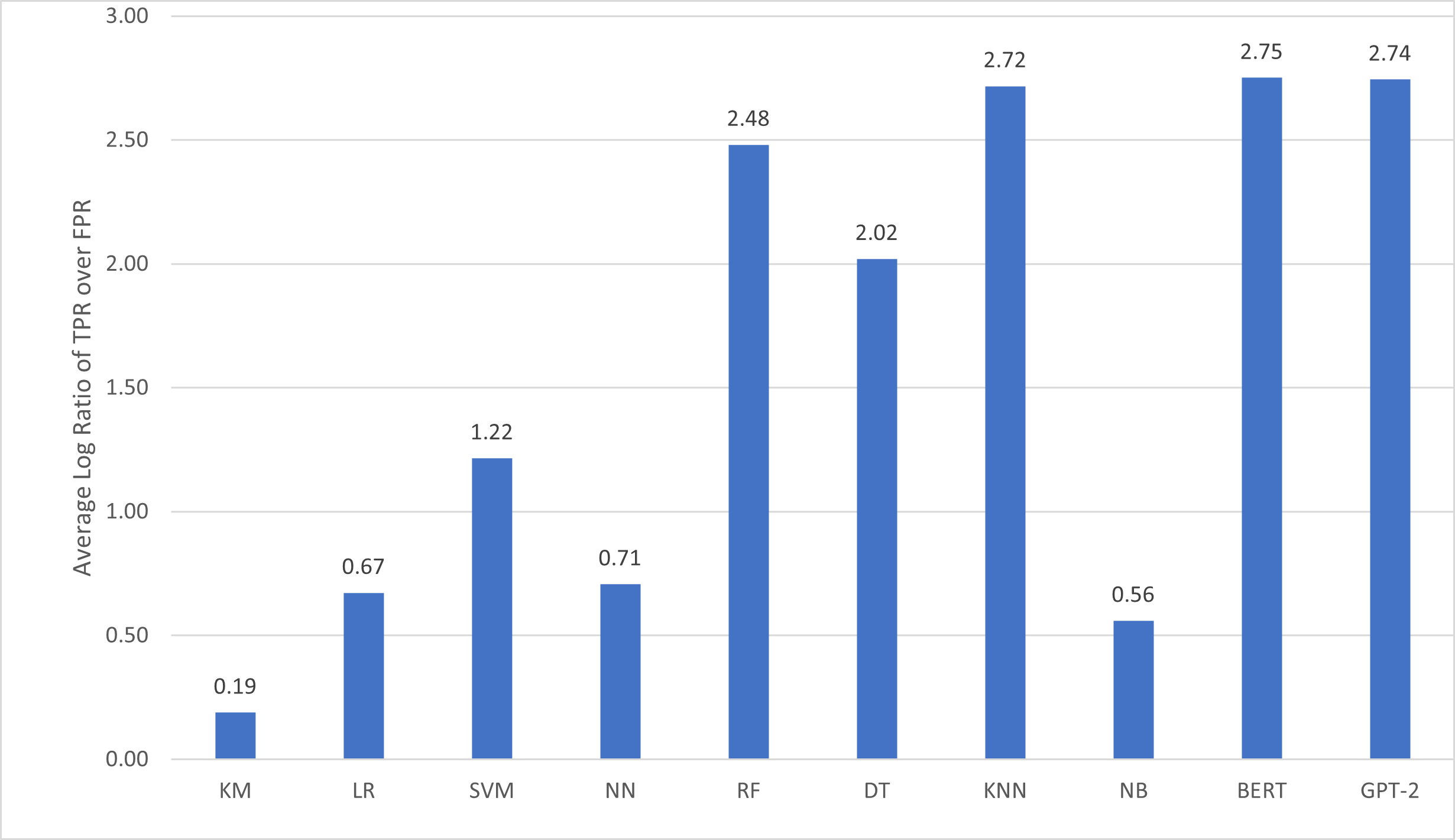}
\caption{Log ratio of recall over false positive rate using different machine learning models}
\label{fig:figure4.3}
\end{figure}

To find out the best algorithm among DT, RF, KNN, BERT, and GPT-2, we take the average of log ratios between recall (TPR) and FPR from Table \ref{table:overall-results} and demonstrate it in Figure \ref{fig:figure4.3}. The highest bar shows the best performing model. Therefore, BERT is the best model with the highest average of log ratio between recall and false positive, which is 2.75. This indicates that BERT yields the highest true positive (recall) at a very low FPR, which is our primary goal in an intrusion detection system. BERT achieves a 0.00 FPR on nine out of 11 datasets (except the Synthetic Lpr and ADFA-LD ones). The model's FPR on the Xlock data is 0.01 and on the ADFA-LD is 0.05. BERT also achieves the highest recalls on all datasets. Additionally, GPT-2 is the second best model with the second highest average of log ratio, which is 2.74. GPT-2 achieves the highest recalls on ten out of 11, and its FPR is always below 0.06. Furthermore, KNN is the third best model with 2.72 average of log ratio. This is because KNN has higher FPR than DT and RF on the Synthetic Ftp, Xlock, and ADFA-LD datasets. On the other hand, K-means is the lowest performing model with the average log ratio of 0.19. NB is the second lowest performing model, whose average log ratio is 0.56. This indicates that these models cannot distinguish between normal sequences and intrusion sequences. Since a low FPR is more important than recall and accuracy, we conclude that BERT is the best candidate in detecting intrusions.

\subsection{Results discussion and data quality assessment}

\begin{figure}[htbp]
\centering

    \includegraphics[width=\linewidth]{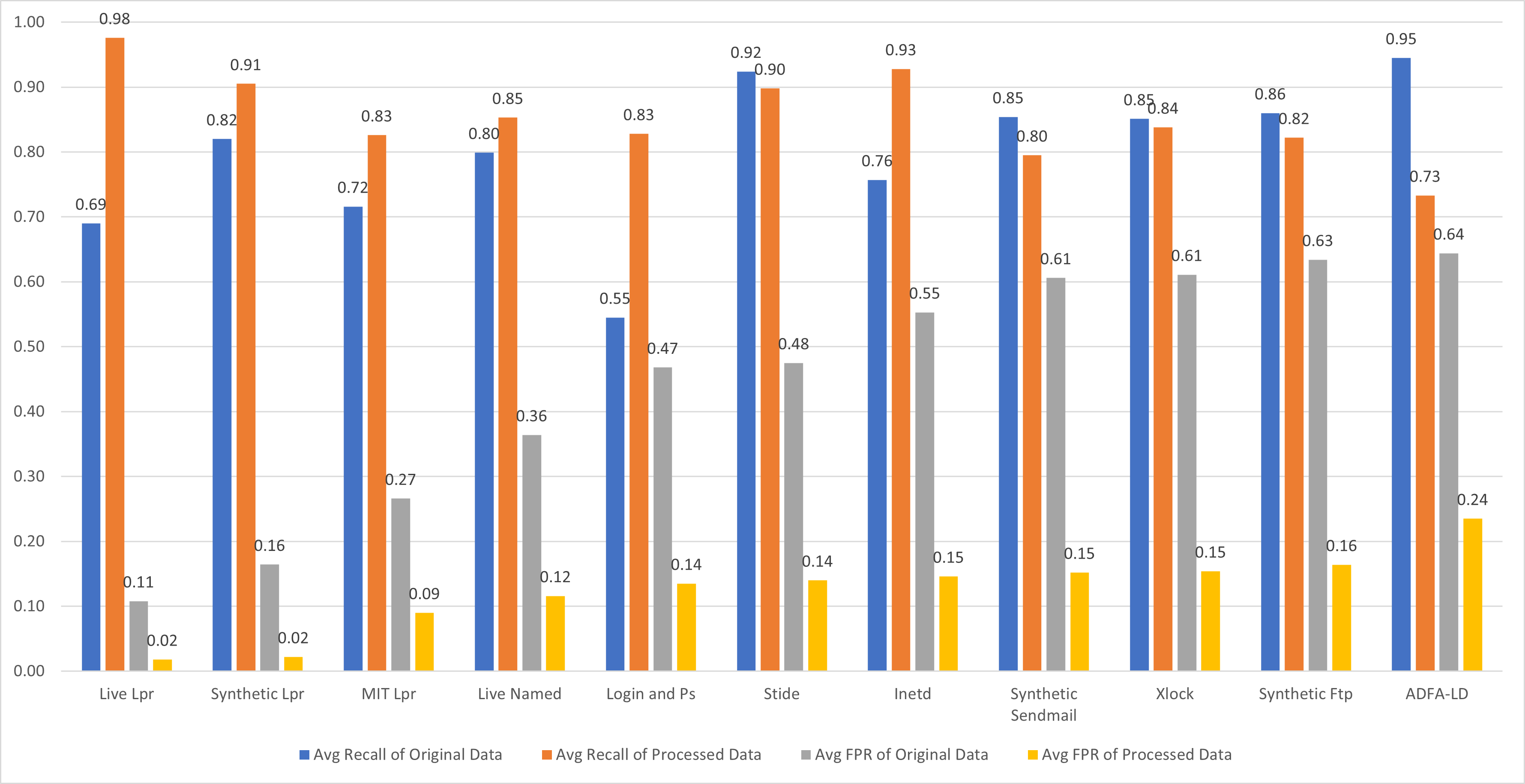}
\caption{A Clustered Bar Chart of Average Recalls and Average FPR from Unprocessed and Processed Data}
\label{figure4.4}
\end{figure}

To compare the overall performance on different datasets, we have created a clustered bar chart of average recalls (TPR) and average FPR - Figure \ref{figure4.4}. This figure is plotted using the performance metrics of processed data and original data, and it is sorted decreasingly by the average FPR. The blue bars are the average recalls of the original data, while the orange bars are the average recalls of the processed data. The grey bars are the average FPR of the original data, whereas the yellow bars are the average of the processed data. By visualizing the difference in performances among the datasets, we can identify the best dataset as well as the worst dataset. After being processed, the Live Lpr dataset yields the highest TPR and the lowest FPR. Likewise, the Synthetic Lpr yields the second highest TPR and the second lowest FPR. Therefore, for HIDS, the Live Lpr is the best dataset and the Synthetic Lpr is the second best. 

Additionally, Figure \ref{figure4.4} is also helpful in delineating the difference in performances before and after a dataset is processed. Based on this figure, the best performing processed datasets with the lowest average FPR are the Live Lpr, Synthetic Lpr, MIT Lpr, and Live Named in decreasing order. This indicates that most candidate algorithms yield very low FPR on these datasets. This is also confirmed by the performance metrics from Table \ref{table:overall-results}. The magnitude of the grey bars indicates that these datasets do not yield such low FPR before being processed. On average, the original Live Lpr dataset has lower quality and, therefore, yields a FPR 20 times higher than the processed one. Similarly, the original Synthetic Lpr yields a FPR 27.5 times higher than the processed one, the original Live Named yields a FPR 5 times higher, and the original MIT Lpr yields a FPR 1.8 times higher. Furthermore, these datasets yield higher recall after being processed. On average, after being processed, the Live Lpr  yields recall 1.22 times higher and the Synthetic Lpr yields recall almost 1.5 times higher. The processed MIT Lpr only increases its average recall by 0.006 because there is no duplication in this dataset. Therefore, the data cleaning process has no effect on this data; hence, its performance was not increased. Although the processed Live Named has lower recall than the original data, its FPR is significantly decreased from 0.644 to 0.116. Therefore, it is safe to say that effective data cleaning has lowered the FPR and increased the TPR and, therefore, improved the models' performance on these datasets. On the other hand, the ADFA-LD dataset has the lowest average recall and the highest FPR. Therefore, it is the worst dataset for HIDS.

Based on these observations, we also conduct data quality assessment to determine the best characteristics that a HIDS dataset should possess in order to yield the best possible result. Using data quality principles mentioned in section \ref{sec:dataqualitydimension}, we have created a data quality evaluation table (Table \ref{table:mlmodels}) to show the detailed characteristics of each dataset. 

As can be seen from Table \ref{table:mlmodels}, Live Lpr, Live Named, and Synthetic Lpr are the three best datasets with the highest performance. Overall, these datasets are very similar to the other datasets in terms of data quality principles and dimensions. However, what set them apart is some of their data characteristics. Live Lpr, Live Named, and Synthetic Lpr have a similarity in data reputation, where data were collected over at least a month. This provides sufficient data from both classes for training and testing purposes. 

Another common ground these datasets have is data comprehensiveness, where original and processed data are imbalanced. The disproportion between normal and intrusion classes is typical and inevitable in a HIDS dataset as intrusion rarely happens. Nevertheless, the outstanding performances on these three datasets have proven that class imbalance is not an issue in HIDS performance since we use the bootstrapping technique as a countermeasure. 

Furthermore, these datasets contain consistent data, where each trace is collected sequentially and then grouped by its PID during data processing. As processes were running concurrently during data collection, this step ensures that each system call trace belongs to only one PID without any interruption from other PIDs. ADFA-LD contains data from different processes without PID, so we could not process them similarly to the UNM and MIT datasets. As traces of system calls are not grouped by their corresponding PID, signature sequences are not guaranteed to be from the same PID. This is a probable reason why ADFA-LD yields the lowest overall performance (0.61). 

Besides, both classes from these datasets are collected and stored separately. This guarantees correct data labeling (data accuracy) and, hence, increases the models' performances. Lastly, having a medium to large overlap percentage between normal sequences and intrusion sequences could have been an obstacle in yielding optimal performances. However, effective data cleaning has helped us avoid this problem. By removing duplicated sequences existing in both classes, we only train the models with unique signature sequences from each class. As a result, this increases recalls, reduces false positives and false negatives, and, therefore, boosts the model performances on these datasets. 

Our analysis shows that the most important qualities in a HIDS dataset are data reputation, data accuracy, and data consistency. Data reputation ensures data sufficiency and trustworthiness in the data source. To achieve this quality, data should be collected over at least a month of activities by a credible institute. Data accuracy guarantees correct data labeling and reliable detection results. This can be achieved by following the same data collection technique from UNM, MIT, and ADFA that is collecting data from different classes separately. Data consistency ensures signature sequences are pure and corresponding to the same PID. This quality can be achieved by recording each system call along with their corresponding PID - similar to the UNM data format. 

\begin{center}
\begin{longtable}{p{1.4cm}p{12cm}p{2cm}}
\caption{\centering Data quality evaluation of the datasets used in the experiment regarding the data quality dimensions discussed in Section \ref{sec:dataqualitydimension}. A: Reputation; B: Relevance; C: Comprehensiveness; D: Timeliness; E: Variety; F: Accuracy; G: Consistency; H: Duplication. Performance represents the average log ratio of TPR over FPR per dataset.} \\
\hline
\textbf{Dataset}& \textbf{Data quality evaluation}& \textbf{Performance}\\
\hline
\multirow{10}{4em}{Synthetic Sendmail} & \textbf{A:} Collected from the sendmail program using strace on Sun SPARC stations running unpatched SunOS 4.1.1 and 4.1.4. \newline \textbf{B:} To monitor normal usage and detect sunsendmailcp intrusion, decode intrusion, and forwarding loops error. \newline \textbf{C:} Both original data (1.8 million normal vs. 6,755 intrusion system calls) and processed data (7,759 normal vs. 451 intrusion sequences) are imbalanced. \newline \textbf{D:} Collected in 1996. \newline \textbf{E:} The signature of both classes covers system call numbers from 1 to 168. \newline \textbf{F:} Data is correctly labeled. Normal and intrusion data are stored in separate folders. \newline \textbf{G:} Data is consistent. Features are extracted sequentially by PIDs. \newline \textbf{H:} There is 99.5\% of overlap between normal sequences and intrusion sequences. & 1.40 \\
\hline
\multirow{10}{4em}{Synthetic Ftp} & \textbf{A:} Collected from the Washington University ftpd server using strace on a Linux machine. \newline \textbf{B:} To monitor normal usage and detect misconfiguration vulnerability. \newline \textbf{C:} Both original data (180,315 normal vs. 1,363 intrusion system calls) and processed data (28,415 normal vs. 376 intrusion sequences) are extremely imbalanced. \newline \textbf{D:} Collected in 1998. \newline \textbf{E:} The signature of both classes covers system call numbers from 1 to 164. \newline \textbf{F:} Data is correctly labeled. Normal and intrusion data are stored in separate folders. \newline \textbf{G:} Data is consistent. Features are extracted sequentially by PIDs. \newline \textbf{H:} There is 84.15\% of overlap between normal sequences and intrusion sequences.  & 1.40 \\
\hline
\multirow{10}{4em}{Synthetic Lpr} & \textbf{A:} Collected from the lpr program over 15 months using strace on Sun SPARC stations running unpatched SunOS 4.1.4. \newline \textbf{B:} To monitor normal usage data and detect lprcp intrusion signature. \newline \textbf{C:} Original data is extremely imbalanced (2,400 normal vs. 164,232 intrusion system calls). Processed data is imbalanced (975 normal vs. 2,232 intrusion sequences). \newline \textbf{D:} Collected in 1991. \newline \textbf{E:} The signature of both classes covers system call numbers from 2 to 168. \newline \textbf{F:} Data is correctly labeled. Normal and intrusion data are stored in separate folders. \newline \textbf{G:} Data is consistent. Features are extracted sequentially by PIDs.  \newline \textbf{H:} There is 98.08\% of overlap between normal sequences and intrusion sequences. & 1.94 \\
\hline
\multirow{10}{4em}{Live Lpr} & \textbf{A:} Live data were collected over 3 months from a SunOS 4.1.4 machine at UNM. \newline \textbf{B:} To monitor normal usage data and lprcp intrusion signature from the same MIT Lpr scripted attack. \newline \textbf{C:} Both original data (187,102 normal vs. 164,232 intrusion system calls) and processed data (108,700 normal vs. 4,000 intrusion sequences) are balanced. \newline \textbf{D:} Collected in 1996. \newline \textbf{E:} The signature of both classes covers system call numbers from 1 to 168. \newline \textbf{F:} Data is correctly labeled. Normal and intrusion data are stored in separate folders. \newline \textbf{G:} Data is consistent. Features are extracted sequentially by PIDs.  \newline \textbf{H:} There is 67.92\% of overlap between normal sequences and intrusion sequences. & 2.34\\
\hline
\multirow{10}{4em}{MIT Live Lpr} & \textbf{A:} Live data were collected over 2 weeks from 77 hosts on SunOS 4.1.4 machines at the MIT lab. \newline \textbf{B:} To monitor normal usage data and lprcp intrusion signatures from scripted attacks. \newline \textbf{C:} Original data is balanced (174,260 normal vs. 165,248 intrusion system calls). \newline \textbf{D:} Collected in 1997. \newline \textbf{E:} The signature of both classes covers system call numbers from 0 to 169. \newline \textbf{F:} Data is correctly labeled. Normal and intrusion data are stored in separate folders. \newline \textbf{G:} Data is consistent. Features are extracted sequentially by PIDs.  \newline \textbf{H:} No duplication. & 1.68\\
\hline
\multirow{10}{4em}{Xlock} & \textbf{A:} Both live and synthetic data from Xlock were collected on a Linux machine over 2 days. \newline \textbf{B:} To monitor normal usage of xlock command and detect a buffer overflow exploit signature. \newline \textbf{C:} Original data is extremely imbalanced (339,177 normal vs. 949 intrusion system calls). Therefore, we only use 25,000 normal system calls. Processed data is imbalanced (19,487 normal vs. 635 intrusion sequences). \newline \textbf{D:} Collected in 1997. \newline \textbf{E:} The signature of both classes covers system call numbers from 1 to 164. \newline \textbf{F:} Data is correctly labeled. Normal and intrusion data are stored in separate folders. \newline \textbf{G:} Data is consistent. Features are extracted sequentially by PIDs.  \newline \textbf{H:} There is 23.09\% of overlap between normal sequences and intrusion sequences. & 1.22 \\
\hline
\multirow{10}{4em}{Live Named} & \textbf{A:} Live normal data was collected over a month from the Named program on a UNM Linux 2.0.35 kernel. \newline \textbf{B:} To monitor normal usage data and detect a buffer overflow exploit. \newline \textbf{C:} Original data is extremely imbalanced (9.2 million normal vs. 1,800 intrusion system calls). Therefore, we only use 2,000 normal system calls to create a more balanced dataset. Processed data is imbalanced (99 normal vs. 273 intrusion sequences). \newline \textbf{D:} Collected in 1998. \newline \textbf{E:} The signature of both classes covers system call numbers from 1 to 141. \newline \textbf{F:} Data is correctly labeled. Normal and intrusion data are stored in separate folders. \newline \textbf{G:} Data is consistent. Features are extracted sequentially by PIDs.  \newline \textbf{H:} There is 90.18\% of overlap between normal sequences and intrusion sequences. & 1.98 \\
\hline
\multirow{10}{4em}{Login and ps} & \textbf{A:} Both live and synthetic data were collected on a 2.0.35 Linux kernel over a month. The  Login version is  from Red Hat util-linux-2.5.38. The Ps version is from Red Hat procps v.1.01. \newline \textbf{B:} To monitor normal usage data and detect Trojan intrusion signature. \newline \textbf{C:} Original data is balanced (15,050 normal vs. 11,825 intrusion system calls). Processed data is imbalanced (176 normal vs. 714 intrusion sequences). \newline \textbf{D:} Collected in 1998. \newline \textbf{E:} The signature of both classes covers system call numbers from 1 to 142. \newline \textbf{F:} Data is correctly labeled. Normal and intrusion data are stored in separate folders. \newline \textbf{G:} Data is consistent. Features are extracted sequentially by PIDs.  \newline \textbf{H:} There is 96.69\% of overlap between normal sequences and intrusion sequences. & 1.73 \\
\hline
\multirow{10}{4em}{Inetd} & \textbf{A:} Live data was collected from the Inetd program on a Linux 2.0.35 kernel at UNM. \newline \textbf{B:} To monitor normal usage data and detect the signature of a DoS attack which ties up all network connection resources. \newline \textbf{C:} Original data is imbalanced (541 normal vs. 8,371 intrusion system calls). Processed data is balanced (536 normal vs. 487 intrusion sequences). \newline \textbf{D:} Collected in 1999. \newline \textbf{E:} The signature of both classes covers system call numbers from 1 to 137. \newline \textbf{F:} Data is correctly labeled. Normal and intrusion data are stored in separate folders. \newline \textbf{G:} Data is consistent. Features are extracted sequentially by PIDs.  \newline \textbf{H:} There is 94.53\% of overlap between two classes, and all normal data appear in the intrusion data. Therefore, we only remove the overlapped intrusion sequences. & 1.70 \\
\hline
\multirow{10}{4em}{Stide} & \textbf{A:} Live Stide data was collected from a modified Linux 2.0.35 kernel at UNM. \newline \textbf{B:} To monitor normal usage data and detect the signature of a DoS attack which affects requesting memory from other programs. \newline \textbf{C:} Original data is extremely unbalanced (15.6 million normal vs. 206,000 intrusion system calls). Therefore, we only used 1.1 million normal system calls. Processed data is imbalanced (17,182 normal vs. 1,562 intrusion sequences). \newline \textbf{D:} Collected in 1999. \newline \textbf{E:} The signature of both classes covers system call numbers from 1 to 136. \newline \textbf{F:} Data is correctly labeled. Normal and intrusion data are stored in separate folders. \newline \textbf{G:} Data is consistent. Features are extracted sequentially by PIDs. \newline \textbf{H:} There is 98.57\% of overlap between normal sequences and intrusion sequences.  & 1.75 \\
\hline
\multirow{10}{4em}{ADFA-LD} & \textbf{A:} Live data was collected from an UbuntuOS version 11.04. \newline \textbf{B:} To monitor normal usage data and detect different types of attack signatures. There are six attack types included in the testing set: brute force attack over open FTP ports and SSH ports, unauthorized root user creation, target host compromise through Java and Linux meterpreter payloads, privilege escalation over webshell. \newline \textbf{C:} Both original data (308,077 normal vs. 317,388 intrusion system calls) and processed data are balanced (161,400 normal vs. 194,000 intrusion sequences). \newline \textbf{D:} Collected in 2013. \newline \textbf{E:} The signature of both classes covers system call numbers from 1 to 325 in Linux kernel 2.6.38. \newline \textbf{F:} Data is correctly labeled. The training and validating sets only contain normal data, and the testing set only contains intrusion data. \newline \textbf{G:} Data is consistent. Features are extracted by specific processes.  \newline \textbf{H:} There is 43.16\% of overlap between normal sequences and intrusion sequences.  & 0.61\\
\hline

\label{table:mlmodels}
\end{longtable}
\end{center}

\section{Conclusion and future work}
Both the quality of a dataset and the capability of a model could contribute to the performance of a machine learning system. However, researchers usually under-value data work vis-a-vis model development \cite{10.1145/3411764.3445518}. In this article, we first discussed the data preparation workflow and data quality attributes for intrusion detection. Taking a HIDS as a case study, we then conducted experiments on 11 datasets using seven machine learning models and three deep learning models. Based on the experimental results, we propose the following conclusions: (1) Deep learning models, such as BERT and GPT-2, outperform the traditional machine learning models for intrusion detection since the former can encode the contextual information for sequential data (Figure \ref{fig:figure4.3}). (2) Almost all of the algorithms achieved a better performance on the Synthetic Lpr, Live Lpr, Xlock, Live Named, Inetd, and Stide datasets than the others, indicating that the data quality of these datasets might be higher than the other datasets (Table \ref{table:overall-results}). (3) Improving the data quality can enhance the performance of the machine learning performance in most of the situations (Figure \ref{figure4.4}). (4) The class imbalance issue in intrusion detection could be solved by using the bootstrapping technique to generate a balanced dataset in different categories. (5) Reputation, accuracy, and consistency are the data quality dimensions which yield high quality datasets for HIDS in this research.

To assure data quality, we need to carefully check the data quality first. Tools such as the data validation system \cite{polyzotis2019data}, BoostClean \cite{krishnan2017boostclean}, and ActiveClean \cite{krishnan2016activeclean} can be used to detect and fix some potential data issues. However, these tools fail to connect data quality with machine learning performance. Therefore, it is necessary to conduct quantitative studies to verify the correlations between data quality and machine learning performance. The second step is to improve the data quality. Different approaches can be used to improve data quality regarding different data quality dimensions. For example, to improve the correctness of a dataset, we may need to remove the label noises. To improve the variety of the dataset, we increase the unique data items in a dataset and make sure the distribution of the dataset follows the distribution of the population. To alleviate the data imbalance issue, we can use the bootstrapping technique to generate a more balanced dataset. Techniques such as transfer learning \cite{9282280} and knowledge graph \cite{bhatt2020knowledge} have also been proved useful for data quality improvement.  

In the future, we will use external resources, such as knowledge graphs, to enhance the semantic representation of the input data to improve the model performance for intrusion detection \cite{bhatt2020knowledge}. More importantly, transfer learning over knowledge graph is also a promising strategy to incorporate domain knowledge for performance improvement for intrusion detection.   

\section*{Acknowledgements}

This work was supported in part by the National Science Foundation under Grant 1852249, in part by the National Security Agency under Grant H98230-20-1-0417. The authors would like to thank Mr. Abdullah Gadi for conducting the investigation of existing public datasets for IDS and creating tables \ref{table:idsdatasets}. The authors are grateful to Ms. Marie Bloechle for editing the paper.

\bibliographystyle{unsrt}  
\bibliography{references}  

\appendix
\section{Appendix A: Abbreviations of the machine learning models}
\label{Appendix:A}

Logistic regression (LR), support vector machine (SVM), decision tree (DT), random forest (RF), neural network (NN), artificial neural network (ANN), isolation forest (IF), k nearest neighbor (KNN), scaled convex hull (SCH), histogram-based outlier score (HBOS), cluster-based local outlier factor (CBLOF), naïve Bayes (NB), averaged one dependence estimator (AODE), radial basis function network (RBFN), multi-layer perceptron (MLP), deep neural networks (DNN), convolutional neural network (CNN), long short-term memory (LSTM), recurrent neural network (RNN), gated recurrent unit (GRU), softmax regression (SR), hidden Markov model (HMM), variational autoencoder (VAE), sequence covering for intrusion detection (SC4ID), intrusion detection tree (IntruDTree).

\end{document}